# TOWARDS DEVISING A FUND MANAGEMENT SYSTEM USING BLOCKCHAIN

Nibula Bente Rashid, Joyeeta Saha, Raonak Islam Prova,
Nowshin Tasfia, Md. Nazrul Huda Shanto and Jannatun Noor,

School of Data and Sciences, BRAC University, Dhaka, Bangladesh

## ABSTRACT

*State government operations comprise a large number of transactions for different processes that must be carried out across the state. This comprises new projects, maintenance and repairs, public employee compensation, and agricultural schemes. Low-level corruption, which is sometimes difficult to trace and hinders state growth, is a big challenge for the top administration. In order to eradicate corruption and bring transparency, technology can be used in an efficient way. An important task to exterminate corruption is to keep track of all the financial transactions of an undergoing project. This research uses blockchain technology to keep track of fund management systems and assure the transparency of any financial statement. This paper proposes to use a gateway where all transaction records are updated in the system and visible to all stakeholders. We find research gaps in the literature and focus on including government funds and local currency usage. The proposed model's motive is to generate a funding model that attains two sub-goals: designing a fund management methodology in which authorized individuals can receive and withdraw allocated funds in crypto currency, and evaluating a smart contract to incorporate the money and identify transparency and tracking. The proposed model executes every feature of our system in just 8.3786ms on average.*

## KEYWORDS

*Blockchain, Ethereum, Smart contract, Government Funding.*

## 1. INTRODUCTION

World economic system is driven on the basis of faithfulness and loyalty. People choose reliable and profitable institutions and media for investment. However, in developing countries like Bangladesh, due to corruption people have lost their faith in the country's financial system. It obstructs the development of the country and creates economic contortion in different public sectors. According to researchers, 1% corruption reduces 0.72% growth rate and 2% productivity of a country [1].

In order to get rid of corruption and bring transparency, we can use blockchain. An important task to exterminate corruption is to keep track of all the financial transactions of an undergoing project [25][26]. Blockchain provides transparency, security, decentralization, non-corruptibility, immutability, consistency, and speed. These characteristics are essential for a trustworthy fund management system.

In developing countries, corruption mainly takes place in the form of bribery and money embezzlement. Whenever the government or any other organization undertakes a complex project, they create funds. Corrupted people target these funds to achieve their self-gain. Often, it is seen that there are no proper records goo fund was utilized. Moreover, Bangladesh is one of the countries with the highest rates of informal payments in regard to public services (GCR 2015-2016). When getting operational licenses such as an electricity connection, almost 60% of companies intend to make informal payments (ES 2017)[31]. Moreover, Bangladesh has made great progress in a variety of social sectors. On the other hand, corruption, nepotism, malinvestment, and misdirected funds have delayed economic growth and stopped the country from progressing. According to a study conducted by Stockholm University, Bangladesh diverts public resources to unproductive sectors, obstructs the government's ability to apply good policies, and reduces public trust in the government [32]. In Bangladesh, in a report of TIB, it is shown that there is more than 61% misuse of total allocated funds for the implementation of a forest project [2]. In another report, it is shown that about 14.36% to 76.92% corruption and misinformation occur in the climate projects [3].

According to a study conducted by the Stockholm University, corruption in Bangladesh diverts public resources to unproductive sectors, obstructs the government's ability to apply good policies, and reduces public trust in the government [20]. According to an NGO survey on everyday corruption, 66 percent of the population paid bribes to authorities in order to get basic government welfare services [21]. However, there are no existing fund management systems that include government projects. The World Bank has pulled out of a project to build Bangladesh's largest bridge, citing corruption concerns [19]. As a result, Bangladesh's image at the international level is now questionable. If the problem is not solved on an urgent basis, Bangladesh will soon be deprived of financial help from foreign organizations [4].

In this paper, the proposed method is to use blockchain technology for tracking of fund management systems for government, private, non-profit organizations, and also on individual levels. Previous papers on blockchain fund management do not include government and governmental funds together. Moreover, The proposed system uses Ethereum which is a smart contract and digital certification platform that offers Ether crypto currency. Here, payments are cryptographically validated and executed by a network of computers with equality. Furthermore, the proposed method includes the use of local currency to covert to crypto currency in the management system for ease of use. To our knowledge, no previous papers proposed to convert local currency to digital currency.

## 2. RELATED WORK

A fund tracking management system is needed for the authority to decrease the rate of corruption. However, corruption exists in many sectors.

### 2.1. Corruption in Bangladesh

In Bangladesh, corruption mainly takes place in the form of bribery and embezzlement. High officials working in a government or non-government organization are mainly responsible for corruption. Bangladesh receives about 63% of the foreign aid as loan and the rest of the aid (37%) is received as grants. In the economic year 2010-11, Bangladesh received $1721.771 millions in terms of commitment and disbursement. OPEC, ADB, and IDA are the leading aid donor organizations [33]. Though Bangladesh has achieved substantial steps to improve assistance efficiency, doubts persist about who are the main beneficiaries of US $1.5 billion foreign aid that the country gets each year. "Whether foreign assistance helps the nation or not is a tricky subject," said Piash Karim, a sociology and economics professor at BRAC University. "People get a relatively little portion of the entire sum, while a crooked clique of NGOs and government leaders profit. There is no way to provide proof, but there have been claims of wrongdoing in foreign-aid projects," he continued. According to Muhammad from the Jahangirnagar University, fighting corruption is the only solution to the issue. "We should assess the whole aiding process and determine what value we really get since we have been receiving foreign help for quite some time," he added. "The whole assistance system's approach is flawed.

It's time to learn how to use our own resources". If the problem is not solved on an urgent basis, our country will soon be deprived of financial help from foreign organizations [25].

Moreover, in order to transfer funds from donor to receiver, there are several payment services that offer money transferring in different currencies. Bangladesh Electronic Funds Transfer Network (BEFTN) is the first paperless digital interbank money transfer system of Bangladesh, launched in February 2011. As a lean-over checkbook clearing system, it supports both debit and credit transactions. Payroll, domestic and international remittances, welfare payments, bill payments, business dividends, security payments, corporate payments, federal tax payments, and individual payments, all types of credit transfers can be handled through this network. Similarly, it accepts debit transactions such as payments like utility bill, insurance premium, association, EMI, and so on [5].

## 2.2. Fund Management with Blockchain

Since funding procedure includes monetary values, it needs high security and user data privacy. Meanwhile, one of the primary components of blockchain is secure identification and anonymity, which makes blockchain technology the perfect match for funding mechanisms. The blockchain ecosystem comprises ICOs, wallets, and exchanges, which are essential parts of crypto currency transit and administration. On top of these fundamental qualities, blockchain offers infrastructure for the development of D Apps that give a user interface to its clients, in this instance contributors and recipients. Furthermore, other blockchain ecosystem elements, such as distributed ledger and distributed storage, consider making blockchain appropriate for funds collection processes by giving organizations, beneficiaries, donors, and legal authorities more control over the information and promoting data transparency [7]. Hence, blockchain may be used to establish a safe money trail [6].

Formal initiatives to make fundraising systems visible, safe, and traceable are very few, and the majority of platforms are unproductive. There have been several initiatives to model the fund gathering process using blockchain technology. A scholarly study [6], offers its own virtual currency named Charity Coin (CC) and covers the operation of charity fundraising with crypto currency. However, they focus on nonprofit organizations. There is no mention of how their system can be utilized to eradicate corruption in governmental large projects funding. Another research paper [8], has created a Hyperledger-based donation system for NGOs. Charity collection deals with dollars but anything in Hyperledger use, restricts their approach. Their system works with predefined users. However, their work is for charitable organizations. Since there is no specific donor for all the time, it changes over time. As a result, it poses a barrier to their system's usability.

 In another paper [9], the authors discuss the basic mechanism to track charitable donations focusing on blockchain technology. The research work [10], proposed a prototype of a blockchain application for Government fund tracking that helps to track transactions using Hyperledger Composer. However, they do not mention which consensus algorithm they use for system reliability. As they used a permissioned blockchain, if the system is managed only by the government officials and if most of the officials come out dishonest then the system can easily be fooled. In a research paper [11], they propose a transparent tendering system using blockchain where the citizens are able to participate in tendering directly and track all fund transactions by using the combination Hyperledger Fabric and Hyperledger Composer. Because of using private networks, a restriction comes forward for the general citizens to visualize the tendering and funding procedures which creates doubt about the system's reliability. Lastly, In paper [9] The author D.Fiergbor describes the blockchain technology of the fund management system in Ghana.

Here, the author focuses on transparency, accuracy, and accountability among the fund managers. The main reason for that proposal is the security issues of data storage.

## 3. METHODOLOGY

In this research experiment, using blockchain technology, the proposed model is to generate a funding model that attains two goals. First, designing a fund management methodology in which authorized individuals can receive and withdraw allocated funds in cryptocurrency. Secondly, evaluating a smart contract to incorporate the money and identify transparency and traceability.

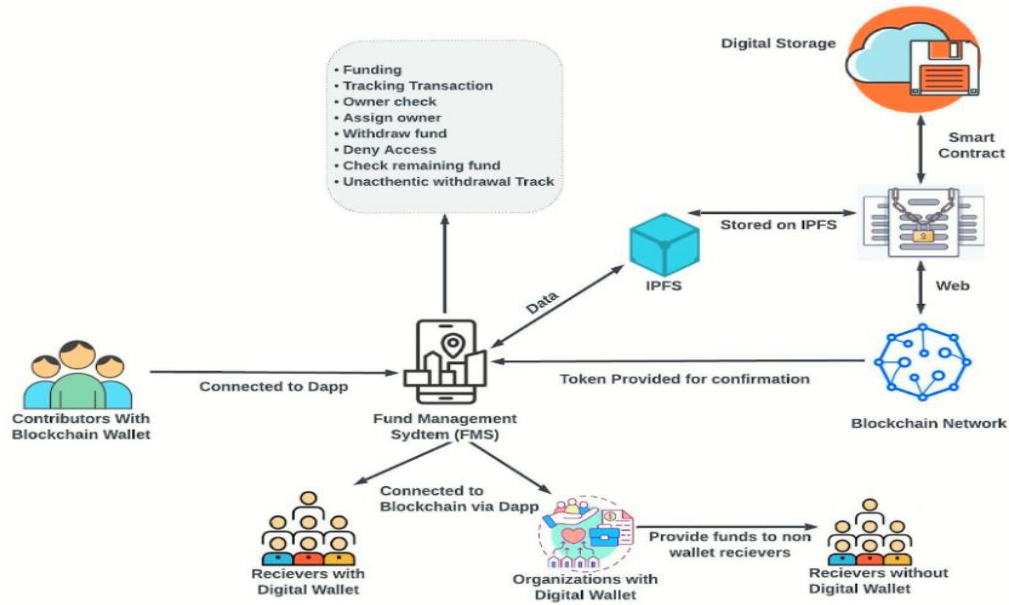

Fig. 1. Framework Architecture

### 3.1. Proposed Framework Architecture

The fund management system can be accessed by the system's three major users: contributors, organizations, and receivers via mobile and web-based applications using blockchain technology. In this system, we use layer 2 scaling mechanism of Ethereum, polygon. As polygon uses different side chains alongside the main chain, it boosts the transactional speed and lowers gas expenses while maintaining the system's decentralization nature as well as security. The proposed framework's high-level structure is shown below in Fig.1 In the proposed model, the person who gives funds in the system is the deployer and he/she is the owner of the transaction. The deployer has full accessibility to track and control the whole transaction system.

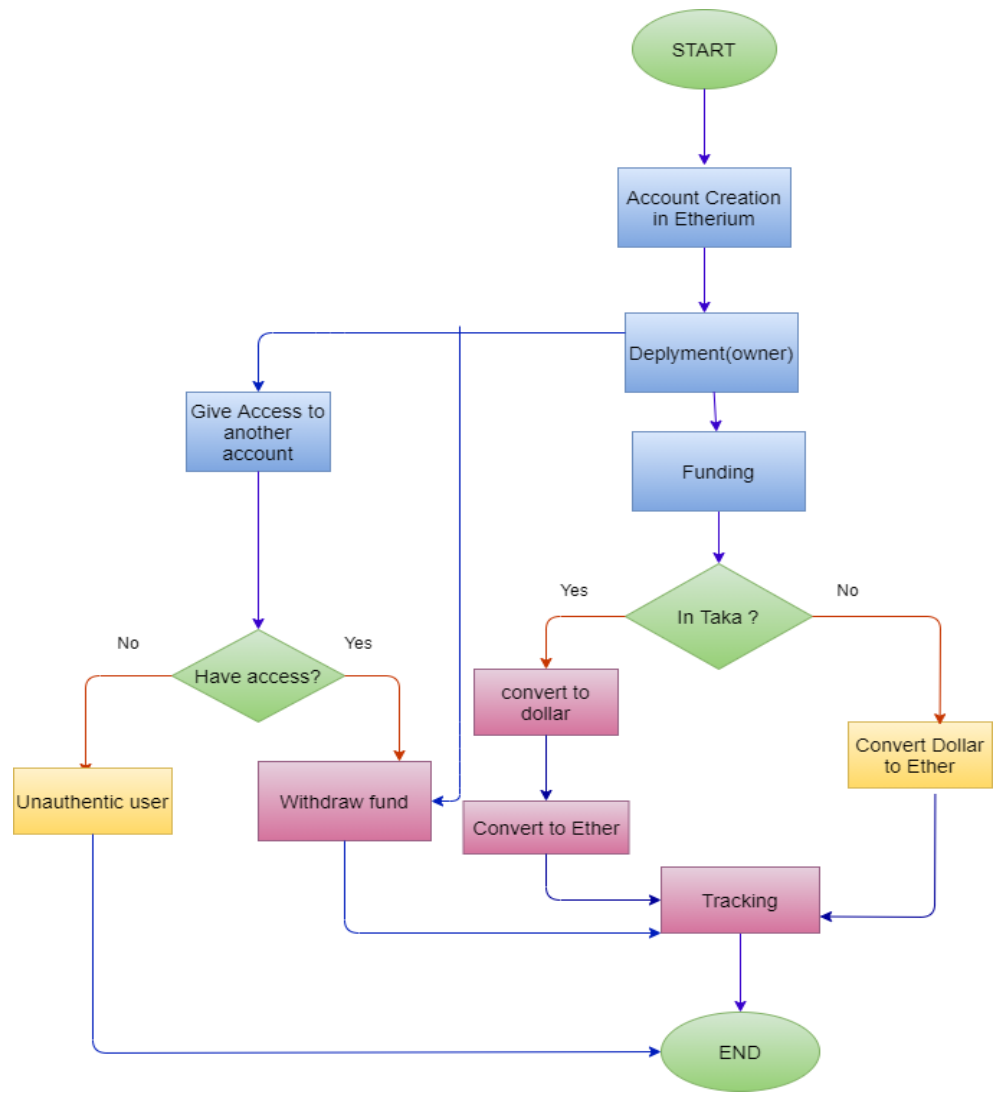

Fig. 2.  Sequential Activity Model

The funder can provide money in local currency through this system. Later, the currency gets converted into crypto currency via the system. Here, only the deployer can withdraw money and also decide whom he or she wants to give access to withdraw money, along with keeping the records of the whole transaction process. The owner has the ability to delete the assigned owner if he or she wants to. If the owner sees any suspicious behavior, then the deployer can delete the suspicious user from the system. Users can check the leftover balance of the funder's account. Users can check and verify the actual owner of the current funding system by the feature. Organizations can also provide documentation by returning images of receipts to ensure that their money is properly distributed among targetted individuals.

## 3.2. Sequential Activity Model of Proposed Model

Fig. 2 shows the system activity model of the system. Here, deployers and authorized users assigned by the owner must register to access the system. The deployer provides access to the user. After gaining access, an authorized individual enters the system using their private key and withdraws funds. Unauthorized users get denied to access the system and get identified as inauthentic users. All other aspects are handled by system deployers, and only authorized people can withdraw funds from the system. After funding, the currency is verified to see if it is in taka(Local Currency) or in dollar using the model. If the currency is in taka, it gets converted to dollars and subsequently to crypto currency. Else, the dollars are turned into crypto currency directly. The whole transaction is recorded and tracked by this blockchain-based system to assure authenticity of the management.

### 3.3. The Layer Structure Of Proposed Model

The presented framework is built on a layered architectural form, as seen in Fig. 3. Interface layer, business logic layer, application layer, transaction layer, trust layer, blockchain layer, security and administrative layer, and infrastructure layer are the eight layers we've created here.

**Interface Layer:** Internet-based websites and DApps of the fund management system are encapsulated in the interface layer. The purpose of this layer is to offer an interface for funders, wallet beneficiaries, and organizations. This layer is used by these individuals to initiate the donation process.

**Business logic layer:** Smart contracts make up the business logic layer, which deals with terms, rules, and interaction requirements. As a result, this layer might be thought of as an operational database of smart contracts, equipped with all interaction, contract activation, and execution rules.

**Application layer:** Online records, donations records, identity verification, and transaction information are all part of the application layer. The application layer combines the interface layers with the business logic as in kind of a smart contract. The application layer combines the interface layers with the business logic as in kind of a smart contract.

**Trust Layer:** The trust layer contains the smart contract's security analysis, formal identity verification, and arbitration methods like Proof-of-Work. The trust layer also interacts with transaction consensus methods and recently introduced block verification, while the blockchain layer stores the results of execution.

**The blockchain layer:** This layer stores data about blocks' and nodes; it also holds the distributed ledger's basic information and hashes of each transaction executed by investors, organizations, and recipients with their secret and public keys addresses.

**Security and administration layer:** The system's security and maintenance is assured by this layer. Several security attacks aim at blockchain, with the 51 percent attack being the most common. This layer is interconnected to the system and works in tandem with it, containing several security algorithms and protocols as well as administrative responsibilities to ensure the system's integrity.

**Transaction layer:** The transaction layer is in charge of transactions, conducted by smart contracts or users of the fund management system.

**Infrastructure Layer:** It is made up of a peer-to-peer network that verifies, forwards, and distributes the Ethereum blockchain transactions. It also covers interaction, authentication, and distributed networking. When one transaction is completed, it is transmitted to all network nodes, and each node verifies the transaction using established parameters. The validated transaction is then saved in the database.

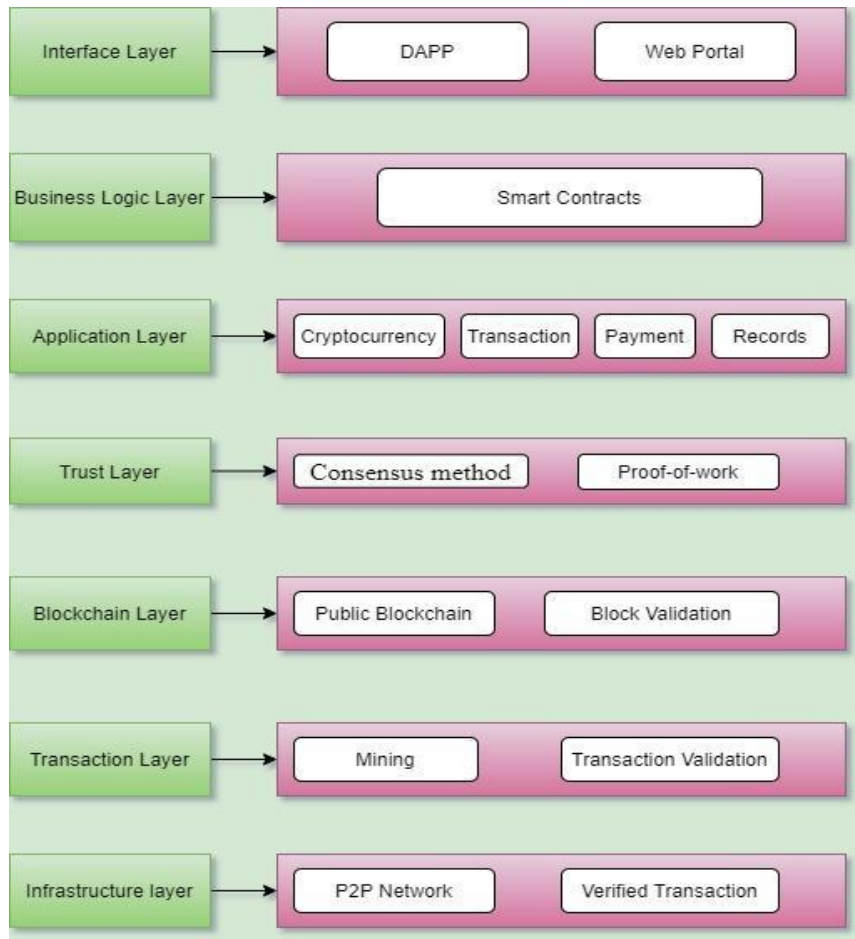

Fig. 3. Layered Structure

## 4. EXPERIMENTAL EVALUATION

Firstly, we discuss the experimental setup and measures of our study. Later, we share some of the results of our proposed system including cost analysis. Lastly, a theoretical comparison of our study and some related studies is shown.

### 4.1. Experimental Setup

For the setup, a metamask extension is integrated into the browser. Metamask is a secure way to connect to blockchain-based applications [13]. We create an account and have to set up polygon network in our metamask. We borrow some matic crypto for our testing purpose. Also, remix IDE is used.

Fig. 4. Fund tracking process

[14]. It is used for smart contract development by using solidity language[15]. Remix IDE is used for testing, debugging, and deploying smart contracts. To execute smart contracts, brownie framework is used, which is a python based development and testing framework [16]. Moreover, to call the solidity language or to execute the whole system we use a brownie framework. In this whole system, we use proof of work to process transactions from peer to peer without any third party[17]. It is a decentralized consensus mechanism. On the other hand, before a transaction, the system goes through the currency conversion. Which means, if the currency is in taka, the system converts it to dollars, and from dollars, it converts to ether. We use oracle [18] for converting currency.

## 4.2. Experimental Findings

In the funding process, the system can track the sender's and receiver's funding information and can also track the withdrawal information as well. In Fig. 4, we see the tracking of transactions after the funding process. Fig. 5 shows the funding of account 2 after the transaction along with the Funding process. Also, Fig. 6 shows the withdrawal of account 1 through which we can check the system's withdrawal feature process. Now, in the proposed model, we calculate response time on the basis of our features. Hence, the execution process in every account remains the same. So, here we differentiate the response time to execute every feature in our system.

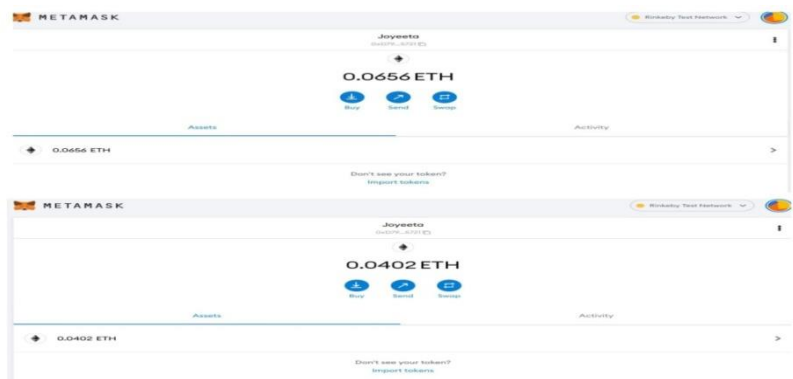

Fig 5.  Funding result of account 2

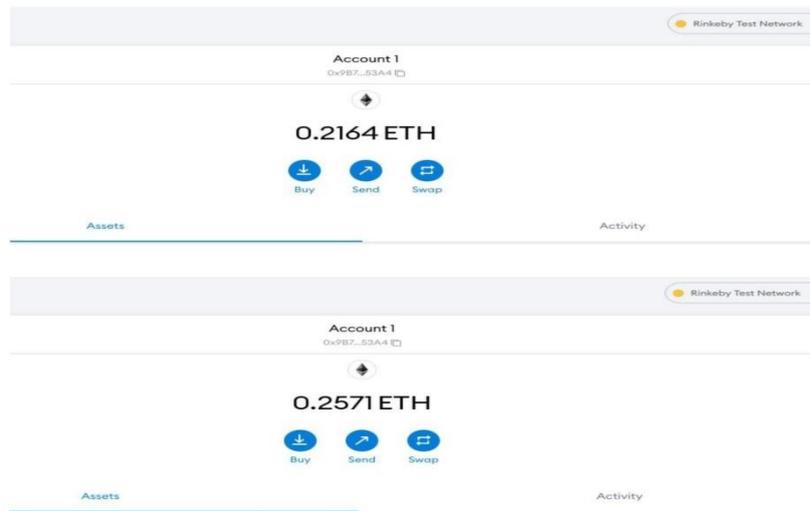

Fig 6. Withdrawal process result of account 1.

Table 1.  Response time for each feature of the proposed model.

| Features | Response time |
|---|---|
| Owner check | 0.3063 ms |
| Owner's balance check | 0.2831 ms |
| Owner's withdrawal process | 6.212270 ms |
| Assign new owner | 39.51796 ms |
| Funding | 0.046349 ms |
| Currency convert in Dollar | 11.9362285 ms |
| Currency Convert in Taka | 11.9585 ms |
| Get Contract Balance | 0.34818 ms |

Table 1 shows the response time for each feature of our proposed model. In our proposed model, it takes 0.304 ms to 40 ms for executing every feature. Hence, on average, to execute the whole system, per account it takes 8.3786 ms including the currency converter in dollars. However, on the basis of currency converter and withdrawal by the assigned owner, the average time may vary. Since the assigned owner may have to go through some process which may cause more time than the withdrawal by the owner itself. The reason for the short response time is the polygon network [12]. We use layer 2 Ethereum scaling of polygon network.

### 4.3.  Cost Analysis and Discussion

Depending on the aforementioned scenarios, we estimate the cost of our proposed model:
-        Procedure
-        A Blockchain App's complexities
-        Development Tools
Costs of deployment and third-party services:

**Public Blockchain:** $0.01 per transaction + $750 for third parties.

**Costs of maintenance:** Approximately 10% to 15% of the total project budget.

The following are some of the third-party tools which Blockchain Applications may require:
Amazon Web Services: Computation, Memory, Delivery. (Depends on the user's number, $100 - \$1000).

**Amazon SNS, Twilio:** Provide notifications inside the interface. (\$10 - \$50).
Mixpanel/Flurry: Analysis of Data, channel, and monitoring. (\$0 - \$150).

The apps can be shifted to multiple platforms depending on their stability, adaptability, and confidentiality since blockchain technology is indeed new to the industry and new systems are arriving in the markets day after day. Hence, further additional costs for development are needed for this system. In our proposed model, we have made a medium complexity blockchain app which is dApps, built on the blockchain platforms like Ethereum. That's why we are able to work with multiple features at a minimal cost.

## 4.4. Theoretical Comparison

As shown in Table 2, related papers do not place government funds as a stakeholder. Also, private and permission less blockchain usage makes our study different than state-of-the-Arts.

Table 2. Theoretical comparison with related works

| Paper Title | Research Gap | Our Proposed Remedy |
|---|---|---|
| A framework to make charity collection transparent and auditable using blockchain technology[6] | No mention of tracking the government funds. | We focus on both governmental and non-governmental organizations |
| E-Governance, A Tendering Framework Using Blockchain[11] | Because of using private network, a restriction comes forward for the general citizens to visualize the tendering and funding procedures which creates confusion on the system's reliability | We use public blockchain in our system. |
| Blockchain for government fund tracking using Hyperledger[10] | As used a permissioned blockchain, if the system is managed by only the government officials and most of the officials comes out as dishonest, then the system can become vulnerable | We use permissionless blockchain in our system |
| A blockchain based tracking system for university donation[22]. | No implementation on how the study materials are going to be tracked | We use Proof of work in our system |
| Blockchain-Based One-Off Address System to Guarantee Transparency and Privacy for a Sustainable Donation | No proper information about how the donors can trust the receivers if their provided donation is properly being | We use decentralized consensus mechanism |

| Environment[23] | utilized or not | |
|---|---|---|
| Platform for Tracking Donations of Charitable Foundations based on Blockchain Technology[24] | Lack of practical implementation, only talked about the tools but not any system prototype. | Implementation Shown |

### 4.5. Potential interdisciplinary applications of this study

Blockchain is not only limited to crypto currency, many other interdisciplinary applications can benefit from Blockchain as well. IoT is a field that is used for solving many social issues. Study [28] works for the safety of elderly populations by monitoring indoor navigational Hazards with the help of IoT. As IoT is vulnerable to security attacks, blockchain can provide security to IoT applications which will make them more reliable. Study [29] discusses how IoT can benefit from blockchain integration.

Study[27], discusses a major issue faced by the elderly populations which is loneliness. To mitigate loneliness government can build infrastructure to create interactive environments for elderly populations to improve their social interaction. In that case, the fund can be tracked by blockchain so that there is no corruption and hence, infrastructure to mitigate loneliness of elderly populations will be ensured. Also, creative industries like music industry can leverage blockchain technology by creating unchangeable database for music copyright information and conducting frictionless royalty payments [30].

## 5. LIMITATIONS AND FUTURE WORKS

A limitation of our study is that we only worked on currency converters of Taka and Dollar. We do not provide options for other currencies. Another limitation is that we use Ethereum which is very costly to handle. As there is a chance that Ethereum will become more expensive to use, that can be a problem since some people may not effort this. On the other hand, in our system, we have a limitation of funding amount. The least amount of fund is 50 dollars. Furthermore, we are now using the proof-of-work mechanism, however, there are other mechanisms that can reduce response time.

In the future, we will try to include different currencies as well. Hence, it would be easier to transfer money or take donations from other countries. Also, in the future, we will try to exclude fund limitations and mention specific withdrawal amount that will give users a more friendly environment. In the future, we will look forward to find a more suitable public blockchain network or try to make one of our own to reduce cost. We can use stable coins instead of Ethereum. Lastly, we will try to shift our mechanism from proof-of-work to proof-of-stake to reduce the response time.

## 6. CONCLUSION

To conclude, in this paper, a blockchain-based fund management system has been introduced. The system can be used in a variety of business settings. The system has taken advantage of blockchain properties like irreversibility and security to create a trustworthy-decentralized system. It outperforms the existing standard methods by implementing a more reliable tracking system which would put an end to any wrongdoings or anomalies. And the smart contract authorities are in charge of making sure that practically all of the system's transactions can be tracked. We use Ethereum as it is a public network which means all the records can be visible and accessible by every peer. Here, by using blockchain, we make a structure for fund management where all the transactional data will be recorded. This data is undeletable and unchangeable. Hence, there is no way to corrupt this structure. With this structure, we can keep records of funds of different sectors so that it would be more trustable to rely on this and will help to reduce corruption as well. Hence, in the government sector or in any organization, this system can be utilized whenever it comes to any trustable fund issues.

## 7. CONCLUSIONS

This paper addresses the lack of strength of OLLVM obfuscation in control flow protection and the gap in identifier obfuscation by proposing two broad categories of enhancements. In control flow obfuscation, first, adding nested switches at the control flow level and adding the switch structure again in the flattened code, thus increasing the complexity of the code while resisting existing scripting attacks; second, proposing an in-degree treatment for bogus blocks to increase the confusion of bogus blocks further. Further, at the level of identifier obfuscation, four algorithms are proposed and bridge the gap of OLLVM in identifier obfuscation. By comparing with OLLVM, this paper can significantly improve the original control flow complexity in obfuscation effect; replace 65.2% of custom identifiers while guaranteeing program functionality. Furthermore, the time overhead from obfuscation is almost negligible. The space overhead is at 1.5 times.

In future work, we will pay attention to generating more secure opaque predicates and are not limited to the number-theoretic model. Meanwhile, the practical effectiveness of existing obfuscation algorithms in large projects remains tested. Therefore, we will focus on how to provide more accessible use of the obfuscation framework model in large projects.

### ACKNOWLEDGEMENTS

*Authors have equal contributions to this paper.

### REFERENCES

[1]  G. Locatelli, G. Mariani, T. Sainati, and M. Greco, "Corruption in public projects and megaprojects: There is an elephant in the room!," International Journal of Project Management, vol. 35, no. 3, pp. 252–268, 2017.

[2]  T. R. . December and T. Report, "61\% fund embezzlement in forest projects: Tib." https://www.tbsnews.net/bangladesh/corruption/ 61-fund-embezzlement-forest-projects-tib-178726, Dec 2020.

[3]  "Tibstudy:14\%to7\%corruption          found          in          climate          change          projects." Https://archive.dhakatribune.com/bangladesh/corruption/2020/12/24/ tib-study-14-to 76- corruption-found-in-climate-change-projects, Dec 2020.

[4]  Bangkok,          "What's          happening          with          aid          to bangladesh?."https://www.thenewhumanitarian.org/report/96902/analysis-what%E2%80%          99s-happening aid bangladesh, Sep 2017.

[5]  "Payment and settlement systems."https://www.bb.org.bd/en/index.php/financialactivity/paysystem..

[6]  M. S. Farooq, M. Khan, and A. Abid, "A framework to make charity collection transparent and auditable using blockchain technology," Computers Electrical Engineering, vol. 83, p. 106588, 2020.

[7]  G. Wood, "Ethereum: a secure decentralized generalized transactio ledger", Ethereum Proj Yellow Pap 2014;151:1–32.

[8]  A. Mehra, S. Lokam, A. Jain, M. Sivathanu, S. Singanamalla, and J. ONeill, "Vishrambh: Trusted philanthropy with end-to-end transparency,"in HCI for Blockchain: a CHI 2018 Workshop on

Studying, Critiquing, Designing and Envisioning Distributed Ledger Technologies, Montreal, QC, Canada, 2018

[9]   D. D. Fiergbor, "Blockchain technology in fund management," in International Conference on Application of Computing and Communication Technologies, pp. 310–319, Springer, 2018

[10]  A. Mohite and A. Acharya, "Blockchain for government fund tracking using hyperledger," in 2018 International Conference on Computational Techniques, Electronics and Mechanical Systems (CTEMS), pp. 231–234, IEEE, 2018

[11]  Y. Goswami, A. Agrawal, and A. Bhatia, "E-governance: A tendering framework using blockchain with active participation of citizens," in 2020 IEEE International Conference on Advancved Networks and Telecommunications Systems (ANTS), pp. 1–4, IEEE, 2020

[12]  S. Neiwal, "Polygon (MATIC): The Swiss Army Knife of Ethereum Scaling", https://www.gemini.com/cryptopedia/polygon-crypto-matic-network- dapps-erc20-token, August 2021

[13]  "The crypto wallet for defi, Web3 Dapps and nfts," MetaMask. [Online]. Available: https://metamask.io/. [Accessed: 28-May-2022].

[14]  ReadTheDocs, "Creating and Deploying a Contract," [Online]. Available: https://remix-ide.readthedocs.io/en/latest/createdeploy.html

[15]  J. Frankenfield, "Smart contracts: What you need to know," Investopedia, 24-Mar-2022. [Online]. Available: https://www.investopedia.com/terms/s/smart-contracts.asp. [Accessed: 28-May-2022]

[16]  "How to deploy a smart contract with brownie," QuickNode. [Online]. Available https://www.quicknode.com/guides/web3-sdks/how-to-deploy-a-smart-contract-with-brownie.[Accessed: 28-May-2022].

[17]  J. Frankenfield, "Proof of work (POW)," Investopedia, 12-May-2022. [Online]. Available: https://www.investopedia.com/terms/p/proof-work.asp. [Accessed: 28-May-2022].

[18]  "Oracle Supply Chain amp; Manufacturing 22A – get started," Oracle Help Center, 19- Apr-2022. [Online]. Available: https://docs.oracle.com/en/cloud/saas/supply-chain management/22a/index.html. [Accessed: 28-May-2022].

[19]  "World Bank cancels Bangladesh bridge loan over corruption" BBC News, 30-June-2012. [Online]. Available: https://www.bbc.com/news/world-south-asia-18655846. [Accessed: 19-June-2022].

[20]  "Corruption in Bangladesh Composition" A. Islam, 6-May-2020. [Online]. Available : https://lekhapora.org/corruption-in-bangladesh-composition/. [Accessed: 19-June-2022].

[21]  B. Report, "Bangladesh ranks 13th most corrupt country in the world." https://www.businessinsiderbd.com/national/news/16468/ bangladesh-ranks-13th-most-corrupt-country-in-the-world, Jan 2022.

[22]  E. A. FERWANA, A BLOCKCHAIN-BASED TRACKING SYSTEM FOR UNIVERSITY DONATION. PhD thesis, 2021.

[23]  J. Lee, A. Seo, Y. Kim, and J. Jeong, "Blockchain-based one-off address system to guarantee transparency and privacy for a sustainable donation environment," Sustainability, vol. 10, no. 12, p. 4422, 2018.

[24]  H. Saleh, S. Avdoshin, and A. Dzhonov, "Platform for tracking donations of charitable foundations based on blockchain technology," in 2019 Actual Prob-lems of Systems and Software Engineering (APSSE), pp. 182–187, IEEE, 2019.

[25]  Benjamin A. Olken, "Monitoring Corruption: Evidence from a Field Experiment in Indonesia," Journal of Political Economy 115, no. 2 (April 2007): 200-249. doi: 10.1086/517935

[26]  ONE Campaign, 2022. Public monitoring of government projects reduced corruption by 20\%.Retrieved July 15, 2022, from https://www.one.org/international/follow-the-money/public-monitoring-of-government-projects-reduced-corruption-by-20/

[27]  Thakur, N., Han, C.Y.: A framework for facilitating human-human interactions to mitigate loneliness in elderly. In: Ahram, T., Taiar, R., Langlois, K., Choplin, A. (eds.) IHIET 2020. AISC, vol. 1253, pp. 322–327. Springer, Cham (2021). https://doi.org/10.1007/978-3-030-55307-4\_49

[28]  Thakur, N., Han, C.Y. (2020). A Framework for Monitoring Indoor Navigational Hazards and Safety of Elderly. In: Stephanidis, C., Antona, M., Gao, Q., Zhou, J. (eds) HCI International 2020 – Late Breaking Papers: Universal Access and Inclusive Design. HCII 2020. Lecture Notes in Computer Science(), vol 12426. Springer, Cham. https://doi.org/10.1007/978-3-030-60149-2\_56

[29]  Abdelmaboud A, Ahmed AIA, Abaker M, Eisa TAE, Albasheer H, Ghorashi SA, Karim FK. Blockchain for IoT Applications: Taxonomy, Platforms, Recent Advances, Challenges and Future Research Directions. Electronics. 2022; 11(4):630. https://doi.org/10.3390/electronics11040630

[30] Arcos, L. C. (2018), "The blockchain technology on the music industry", Brazilian Journal of Operations & Production Management, Vol. 15, No. 3, pp. 439-443, doi:10.14488/BJOPM.2018.v15.n3.a11,available from: https://bjopm.emnuvens.com.br/bjopm/article/view/449 (accessed on 13/7/2022).

[31] https://www.ganintegrity.com/portal/country-profiles/bangladesh/, Nov 2020.

[32] A. A. I. a. Arif, "Contact." https://lekhapora.org/ corruption-in-bangladesh-composition/, May 2020.

[33] M. H. Ahamad, "Foreign grants and loans in bangladesh," 2018.

[34] Bangkok, "What's happening with aid to bangladesh?." https: //www.thenewhumanitarian.org/report/96902/analysis-what%E2%80% 99s-happening-aid-bangladesh, Sep 2017.

## AUTHORS

**Nibula Bente Rashid:** recently graduated in computer science and engineering from Brac University, Dhaka, Bangladesh. Her research areas include cloud computing, blockchain, and cybersecurity

**Joyeeta Saha** has recently graduated from Brac University. She was in Computer Science and Engineering Department. Her research interest is related to Blockchain, Cloud computing, and Deep learning. Now she is working as a project employee at Quantanite.

**Raonak Islam Prova** is currently working as a trainee automation developer in Summit14 Communications Limited. She received her B.Sc degree in Computer Science and Engineering from BRAC University. Her fields of interest in research includes cloud computing, blockchain, automation, networking and cybersecurity.

**Nowshin Tasfia** has recently graduated from Brac University. She was in Computer Science and Engineering Department. Her research interest is related to Blockchain, Cloud computing, and Deep learning

**Md. Nazrul Huda Shanto** is working as a Research Assistant at BRAC University. He is currently pursuing a bachelor's degree in Computer Science from the School of Data and Sciences, Brac University, Dhaka, Bangladesh. His area of research interests includes HCI4D, Cloud Computing, Cybersecurity, and Image Processing

**Jannatun Noor** has been working as a Lecturer at the Department of CSE, BRAC University since September 2018. Besides, she is working as a Graduate Research Assistant under the supervision of Prof Dr. A. B. M. Alim Al Islam in the Department of CSE at Bangladesh University of Engineering and Technology (BUET). Prior to BRACU, she worked as a TEAM LEAD of the IPV-Cloud team in IPvision Canada Inc. She received her B.Sc. and M.Sc. degrees in CSE from BUET. Her research work covers Cloud Computing, Wireless Networking, Data Mining, Big Data Analysis, Internet of Things, etc.